\documentclass[review]{elsarticle}

\usepackage{lineno,hyperref}
\usepackage{amsmath}
\usepackage{arydshln}
\usepackage{xcolor}
\usepackage{ulem}
\usepackage{braket}
\journal{Journal of \LaTeX\ Templates}

\begin{document}

\begin{frontmatter}

\title{Electronic isotope shift factors for the Ir $5d^{7}6s^{2} \ ^{4}\!F_{9/2} \to (\mbox{odd},J= 9/2)$ line at 247.587~nm}




\author[ULB,LUND]{Sacha Schiffmann\corref{Sacha}}
\author[ULB]{Michel Godefroid}
\address[ULB]{Spectroscopy, Quantum Chemistry and Atmospheric Remote Sensing (SQUARES), CP160/09, Université libre de Bruxelles (ULB), 1050 Brussels, Belgium}
\address[LUND]{Division of Mathematical Physics, Department of Physics,Lund University, SE-22100 Lund, Sweden}
\cortext[Sacha]{saschiff@ulb.ac.be}

\begin{abstract}
We present the theoretical calculations of the electronic isotope shift factors of the $5d^{7}6s^{2} \ ^{4}\!F_{9/2} \to (\mbox{odd},J= 9/2)$ line at 247.587~nm, that were recently used to extract nuclear mean square radii and nuclear deformations of iridium isotopes~[Mukai~\textit{et al.} (2020)]. The fully relativistic multiconfiguration Dirac-Hartree-Fock method and the relativistic configuration interaction method were used to perform the atomic structure calculations. Additional properties such as the \textit{sharing rule}, Land\'e $g$ factors or \textit{phase tracking} were employed to ensure an adequate description of the targeted odd level. 
\end{abstract}

\begin{keyword}
Isotope shift \sep MCDHF \sep iridium \sep nuclear deformations
\MSC[2020] 00-01\sep  99-00
\end{keyword}

\end{frontmatter}


\section{Introduction}
It is well known that atomic physics can be used as a tool to probe nuclear properties of the atomic nucleus~\cite{Yordanov:20,Wraith2017}. The interactions between the nucleus and the electron cloud induce shifts in the atomic energy levels, which therefore affect each transition line differently. Two kinds of interaction, arising from fundamentally different considerations, relate nuclear and atomic physics. The hyperfine structure results from the electromagnetic interaction between the electromagnetic field of the electron cloud and the nuclear electromagnetic moments~\cite{Arm:71a,LinRos:74a}. The isotope shift describes the energy shift observed for a given line between two different isotopes~\cite{Kin:84a}. We focus here on the latter interaction, that provides information about the change in the mean square radii and nuclear deformations along an isotopic chain~\cite{Papoulia2016}. 

Iridium has a complex, interesting atomic structure as its ground state, [Xe]$4f^{14}5d^{7}6s^{2}$, where [Xe] is a short notation for the xenon core, exhibits a closed $6s$ shell together with an open $d$ shell. Its homologous atoms Rh ($Z=45$) and Co ($Z=27$) provide evidence that the $(nd(n+1)s)^{9}$ ground state atomic configuration oscillates between the $nd^7(n+1)s^2$ and $n'd^8(n'+1)s$ configurations, where $n=3$ and $5$ for Co and Ir, respectively, and $n'=4$ for Rh. Similar observations are  made for the neighbouring Fe-Ru-Os and Ni-Pd-Pt systems
belonging to the  group~VIII of the periodic table. This provides a hint to the reasons why \textit{ab initio} atomic calculations are so scarce for these systems. 

We present the details of the calculations of the isotope shifts parameters recently used by Mukai~\textit{et al.}~\cite{Muketal:2020a} to extract nuclear mean radii and nuclear deformations for $^{193,196,197,198}$Ir isotopes. 
First, we describe the theoretical background in Section~\ref{sec:theory}, then we provide the details of our computational strategy in Section~\ref{sec:computa} and finally Section~\ref{sec:results} is dedicated to the analysis of the results.

\section{Theory}
\label{sec:theory}
\subsection{The MCDHF method}
The calculations were performed using the variational multiconfigurational Dirac-Hartree-Fock (MCDHF) method. The details of the method can be found elsewhere~\cite{Fischer2016,Grant}. We only review here its most important features that are needed to understand our calculations. 

In the fully relativistic framework of the MCDHF approach, the atomic state function (ASF), $\Psi$, of total angular momentum $J$ and parity $\Pi$ is expanded over $jj$-coupled configuration state functions (CSFs), $\Phi$, as
\begin{equation}
    \label{eq:ASF}
    \Psi(\gamma J\Pi) = \sum_{i=1}^{N_{\rm CSF}} c^{\gamma}_i \Phi(\gamma_i J\Pi) \ ,
\end{equation}
where $c^\gamma_i$ are expansion coefficients, $\gamma_i$ unambiguously specifies the angular coupling tree of the $ith$ CSF and $\gamma$ defines the state of interest among all possible states of symmetry $J^\Pi$. The CSFs form an orthonormal basis of functions, which leads to the normalization condition of the ASF, $\sum_{i}^{N_{\rm CSF}} (c^\gamma_i)^2=1$. They are themselves built as antisymmetrized products of one-electron Dirac spinors or relativistic \textit{orbitals}, $\psi_{n  \kappa  m}$, where $n$ is the principal quantum number, $\kappa=\mp j + s$ for $j=l\pm s$ is the relativistic quantum number and $m$ is the magnetic quantum number associated with the $z$-projection of ${\bf j}$. The $j$, $l$ and $s=1/2$ quantum numbers refer to the usual total, angular and spin momentum of a single electron.

The radial parts of the orbitals and the expansion coefficients are obtained by solving iteratively the self-consistent field MCDHF equations derived by applying the variational principle to the energy functional
\begin{equation}
\mathcal{F}[\Psi(\gamma J\Pi) ] = \langle \Psi(\gamma J\Pi)  | {\rm H}_{\rm DC} | \Psi(\gamma J\Pi)   \rangle + \sum_{ab} \
\delta_{\kappa_a \kappa_b} \ \lambda_{ab} \left(\langle \psi_{a} | \psi_{b} \rangle - \delta_{n_a,n_b}\right)  \ ,
\end{equation}
where ${\rm H}_{\rm DC}$ is the Dirac-Coulomb Hamiltonian and the sum over all orbital pairs
$a\equiv n_a \kappa_a \ b\equiv n_b \kappa_b $ 
introduces the orthonormality constraints on the one-electron radial orbitals of the same $\kappa$-symmetry $(\kappa_a = \kappa_b)$ through the Lagrange multipliers $\lambda_{ab}$~\cite{Fischer2016}. When $n_L$ atomic states are simultaneously targeted, the energy functional is built as the weighted sum~\cite{GRANT1980b}
\begin{equation}
    \label{eq:FEOL}
    \mathcal{F}_{\text{EOL}} = \sum_{i=1}^{N_{\rm CSF}}\sum_{j=1}^{N_{\rm CSF}}d_{ij}\mathcal{H}_{ij} + \sum_{ab}
    \delta_{\kappa_a \kappa_b} 
    \lambda_{ab}\ \left(\langle \psi_{a} | \psi_{b} \rangle - \delta_{n_a,n_b}\right)  \,
\end{equation}
where $\mathcal{H}_{ij}$ are the matrix elements $\langle \Phi_i | {\rm H}_{\rm DC} | \Phi_j\rangle$ and the weights $d_{ij}$ depend on the mixing coefficients
\begin{equation}
    \label{eq:d}
    d_{ij}=\sum_{r=1}^{n_{L}}(2J_r + 1)  c_i^{r}c_j^{r}/\sum_{r=1}^{n_{L}}(2J_r + 1) 
\end{equation}
where $c_i^{r}$ and $a_j^{r}$ are the mixing coefficients of the CSFs $i$ and $j$, respectively, in the CSF expansion of the $rth$ ASF. The subscript EOL refers to the \textit{extended optimal level} scheme that allows us to build a single orbital basis optimal for (i) several states of a single $J^\Pi$ symmetry, (ii) states of different parities $\Pi$, (iii) states of different $J$, and (iv) any combination of the cases (i), (ii) and (iii).

Following the MCDHF calculations, relativistic configuration interaction (RCI) calculations are performed without affecting the pre-optimized one-electron orbital basis. Mixing coefficients and energy levels are computed by diagonalizing the Hamiltonian interaction matrix. 
RCI calculations are often used to enhance the description of the ASF obtained from MCDHF calculations, which is mostly dedicated to the optimization of the orbital basis, by increasing the number of CSFs in the ASF expansion and/or by including additional contributions to ${\rm H}_{\rm DC}$, e.g.,

\begin{equation}
\label{eq:DCB+QED}
    {\rm H}_{\rm DCB+QED} = {\rm H}_{\rm DC} + {\rm H}_{\rm B} + {\rm H}_{\rm QED} =  {\rm H}_{\rm DCB} + {\rm H}_{\rm QED}
\end{equation}

\noindent where ${\rm H}_{\rm B}$ is the Breit interaction Hamiltonian in the long wavelength approximation and ${\rm H}_{\rm QED}$ introduces QED operators, such as the self-energy or the vacuum polarization~\cite{Fischer2016}, ultimately building and diagonalizing the interaction matrix in the CSFs space.\\

In this work, we used the {\sc Grasp2018} package~\cite{Fischer:2019aa}  to perform our calculations.

\subsection{Isotope shifts}

The isotope shift (IS) refers to the frequency shift observed for a given transition of wavelength $k$ between two isotopes of mass numbers $A$ and $A'$

\begin{equation}
\label{eq:deltanu}
    \delta\nu_{\text{IS},k}^{AA'} = \nu_{\text{IS},k}^{A} - \nu_{\text{IS},k}^{A'}\ .
\end{equation}
With this convention, a positive $\delta\nu_{\text{IS},k}^{AA'}$ value with $A > A'$  corresponds to a 'normal' isotope shift, referring to the normal mass shift for non-relativistic hydrogenic systems, i.e., a larger frequency for the heavier isotope.
The IS between two given isotopes has two well-known origins: their mass difference and their different nuclear charge density. 
 They are referred to as \textit{mass shift} (MS) and \textit{field shift} (FS), respectively. The MS shift introduces the electronic factor $\Delta K_{\text{MS},k}=K_{\text{MS},u}-K_{\text{MS},l}$ for the transition line $k$ between an upper atomic state $u$ and a lower atomic state $l$, which results from the evaluation of the matrix element of the atomic two-body operators described in Ref.~\cite{Ekman2019}. The frequency mass shift between two isotopes A and A' of mass M and M', respectively, is given by
\begin{equation}
    \delta\nu_{\text{MS},k}^{AA'} = \left(\frac{1}{M} - \frac{1}{M'} \right) \frac{\Delta K_{\text{MS},k}}{h} 
\end{equation}
where $h$ is the Planck constant, so that $\delta\nu_{\text{MS},k}^{AA'}$ is expressed in frequency units. Note that the nuclear mass factor 
$(M'-M)/MM'$, quickly decreasing with $M$,
makes the field shift  predominant in the case of heavy  elements~\cite{BAUCHE197639}.
Since iridium isotopes have large mass numbers ranging from $182$ to $198$, we focus in the present work on the FS.
The latter contribution also couples nuclear and atomic properties. 
In the approximation of a constant electron density within the nuclear volume,  the FS can be evaluated from the expression
\begin{equation}
    \label{eq:F0}
    \delta\nu_{\text{FS},k}^{AA'} \approx \frac{\Delta F_{0,k}}{h}\delta \langle r^{2} \rangle^{AA'}
\end{equation}
as it was done ~\cite{GAMRATH2018,Filippin2017,Filippin2016c,Palmeri2016} using the RIS3 program~\cite{Naze2013}
designed as a module of the {\sc Grasp2K} atomic structure package \cite{Jonsson2007}.

In the MCDHF framework, the field shift has been reformulated recently~\cite{Ekman2019} and coded in a new module, RIS4, fully compatible with the {\sc Grasp} packages~\cite{Jonsson2007,Fischer:2019aa}. As fully detailed in this work, its expression
\begin{equation}
\label{eq:F2i}
    \delta\nu_{\text{FS},k}^{AA'} \approx \sum_{i=0}^{3}\frac{ F_{2i,u}-F_{2i,l}}{h}\ \delta\langle r^{2(i+1)} \rangle^{AA'} = \sum_{i=0}^{3}\frac{\Delta F_{2i,k}}{h}\ \delta\langle r^{2(i+1)} \rangle^{AA'}
\end{equation}
associates the even polynomial expansion of the electron density close to the origin, i.e., within the nuclear volume,
\begin{equation}
    \label{eq:ved}
    \rho^{e}(\mathbf{r}) \approx \sum_{i=0}^{3} \rho_{2i} \ r^{2i} \ ,
\end{equation}
which is truncated to the 6th order, to the difference in nuclear moments $\delta \langle r^2 \rangle^{AA'}=\langle r^2 \rangle^{A}-\langle r^2 \rangle^{A'}$, $\delta\langle r^4 \rangle^{AA'}$, $\delta\langle r^6 \rangle^{AA'}$ and $\delta\langle r^8 \rangle^{AA'}$. 
This approach  permits to go beyond the constant density approximation of Eq.~(\ref{eq:F0}) but 
 requires the knowledge of the higher-order nuclear moments or similarly, an accurate nuclear charge density. The benefit of  the reformulated FS  is the possibility to study nuclear deformations, as demonstrated by Papoulia \textit{et al.}~\cite{Papoulia2016}. As suggested in Ref.~\cite{Ekman2019}, an intermediate solution is to include the effect of a varying electronic density (\textit{ved}) (introduced in Eq.~(\ref{eq:ved})) to evaluate the FS, without considering higher order nuclear moments. Introducing the electronic factors $\Delta F_{0,k}^{(0),ved}$ and $\Delta F_{0,k}
^{(1),ved}$, the FS reduces to

\begin{equation}
    \label{eq:Fved}
    \delta\nu_{\text{FS},k}^{AA'} \approx \frac{1}{h} \left(\Delta F^{(0),ved}_{0,k} + \Delta F^{(1),ved}_{0,k}\ \delta\langle r^{2} \rangle^{AA'} \right)\delta \langle r^{2} \rangle^{AA'} \ ,
\end{equation}
in which only the lowest nuclear moment $\langle r^{2} \rangle$ appears (see Ref.~\cite{Ekman2019} for more details).

\section{The computational strategy}
\label{sec:computa}
We present the calculations for the $5d^7 6s^2 \ ^{4}\!F_{9/2} \to J^\Pi=(9/2)^{\rm o}$ transition line at 247~nm, corresponding to a transition energy of 40~389.83~cm$^{-1}$, for which the IS was recently measured~\cite{Muketal:2020a}. The targeted transition connects the iridium ground state to the sixth odd-parity J=9/2 level that has not yet been assigned with absolute certainty. Both the even and odd parities present different computational difficulties that are detailed in the next sections.
\subsection{Stability of the ground state}
\label{sec:even}
The two lowest identified levels of Ir are both even ${}^{4}\!F_{9/2}$ states, belonging to the $5d^{7}6s^{2}$ and $5d^{8}6s$ configurations, respectively. Their energy difference, $2~834.98$~cm$^{-1}$, is small, and the $5d^{8}6s \ ^{4}\!F_{9/2}$ state is the lowest excited state, considering both parities and all $J$-values~\cite{NIST_ASD,moore1958}. As already pointed out in the introduction, the ground configuration varies along the Ir homogeneous sequence (Co-Rh-Ir). Extra care is therefore required in the atomic structure calculations to ensure that the ground state is correctly described. In a preliminary calculation, we targeted the two lowest even states only, using a simple correlation model, in which substitutions from the  $6s^{2}$ electron pair were allowed to $nln'l'$, where $nl, n'l'$ belong to the orbital active set $\{7s,\ 6p, \ 6d, \ 5f, \ 5g \}$. In addition, the substitutions 
\begin{equation}
5d^{7}6s^{2} \to 5d^{8}6s \quad \mathrm{and} \quad 5d^{7}6s^{2} \to 5d^{9}
\end{equation}
were also included in the CSF expansions. This calculation 
illustrates how much the computational strategy may affect the theoretical energy spectrum: in this model indeed, the $5d^{8}6s \ J=9/2$ level lies 1~094~cm$^{-1}$ below the $5d^{7}6s^{2} \ J=9/2$ level while the latter is the  ground state according to the NIST AEL database~\cite{NIST_ASD}.
Moreover, the eigenvector composition analysis reveals that the lowest $J=9/2$ level is largely dominated by the $5d^{8}6s$ component ($\approx 90 \%$), which excludes the possibility that a large configuration mixing could be alone responsible for a ground state inversion. Results rather suggest a differential sensitivity to electron correlation for the $5d^{7}6s^{2}$ and $5d^{8}6s$ configurations. Separate, simple Dirac-Fock calculations for both configurations result in substantially different $5d$ orbitals, which could explain the different sensitivity of the two configurations. Table~\ref{tab:even_GS} displays the individual energy $\epsilon$ and the mean radius $\langle r \rangle$ for the $5d_{3/2}$, $5d_{5/2}$ and $6s_{1/2}$ orbitals, in both $5d^{7}6s^{2}$ and $5d^{8}6s$ configurations. The two $5d_{3/2}$ and $5d_{5/2}$ orbitals are less bound in the $5d^{8}6s$ configuration as attest their larger mean radius and their lower binding energy, which is consistent with the addition of an extra electron in the $5d$ shell. This observation suggests that an efficient approach could well benefit from the introduction of radial non-orthogonalities~\cite{Fischer1997,Godefroid1997}, using two different sets of non-orthogonal orbitals within each $l_j$-symmetry.
 , i.e. $\{ 5d_{3/2}, 5d_{5/2} \}$ and 
 $\{ 5d'_{3/2}, 5d'_{5/2} \}$ 
for $5d^7 6s^2$ and $5d'^8 6s$, respectively. Unfortunately, such an approach is not allowed with the current version of {\sc Grasp}. Another method will then have to be found to fix this configuration interchange issue, as described in Section~\ref{sec:LSC}.
\begin{table}[!h]
    \centering
        \caption{Individual energy $\epsilon$ and the mean radius $\langle r \rangle$ for the $5d_{3/2}$, $5d_{5/2}$ and $6s_{1/2}$ orbitals, in both $5d^{7}6s^{2}$ and $5d^{8}6s$ configurations}
    \label{tab:even_GS}
    \begin{tabular}{cccccc}
        \hline
        \hline
         & \multicolumn{2}{c}{$\epsilon$ ($E_h$) } && \multicolumn{2}{c}{$\langle r \rangle$ ($a_0$)} \\
         \cline{2-3}
         \cline{5-6}
        $nl_j$ & $5d^{7}6s^{2}$ &$5d^{8}6s$& & $5d^{7}6s^{2}$ &$5d^{8}6s$  \\
        \hline
         $5d_{3/2}$& 0.498 & 0.403 &~~~& 1.63935 & 1.69885 \\
         $5d_{5/2}$& 0.470 & 0.362 &~~~& 1.69950 & 1.79086 \\
         $6s_{1/2}$& 0.296 & 0.293 &~~~& 3.09602 & 3.14508 \\
         \hline
         \hline
    \end{tabular}
\end{table}

\subsection{Sharing rule}
\label{sec:sharing}
The odd-parity states of Ir were investigated in a series of experimental and semi-empirical studies~\cite{moore1958,Kleef1957,Childs1974,XU200752}. Although the few lowest odd-parity $J=9/2$ levels were assigned unambiguously to the $5d^7 6s6p$ configuration, identifications of higher levels, including the presently targeted sixth level, were not considered to be certain. In Ref.~\cite{Kleef1957}, the level at 40~393.83 cm$^{-1}$ is assigned to the $5d^6 6s^26p$ configuration, while is left unassigned in Ref.~\cite{XU200752} and in the NIST ASD database~\cite{NIST_ASD}. Moreover, the semi-empirical calculations from Ref.~\cite{XU200752} suggest that odd-states are highly mixed. In particular, their calculations revealed the following strong configuration mixing $69.5\% \ 5d^76s6p + 23.6\% \ 5d^86p + 6.6\% \ 5d^66s^26p$~\cite{Quinet2020} for the state targeted in the present work. \\

Configuration mixing is known to influence greatly the isotope shift parameters. Using the so-called \textit{sharing rule}~\cite{BAUCHE197639} that omits the off-diagonal contributions arising from single-electron excitations $n \kappa \rightarrow n’ \kappa$, we can separate and add the contributions to the isotope shifts of the $m$ leading configurations that build the ASF. This way, the isotope shift parameters can be used as an interesting tool to reveal the underlying configuration mixing.
This sharing rule was successfully used in experimental studies of isotope shifts in Ir~I~\cite{Sawatzky:1989aa} and Ce~II~\cite{Ishida1997}. Sawatzky and Winkler~\cite{Sawatzky:1989aa} investigated the configuration mixing in the even-parity part of the Ir~I spectrum. Using their measurements for multiple lines simultaneously, they were ultimately able to determine the weights of different configurations for several levels involved in their measurements.  \\

In the case of the field shift parameter $\Delta F_{0,k}$, the sharing rule similarly applies to the change of electron density at the origin as
\begin{equation}
    \Delta \rho_{ul}(0)  \equiv \rho_u(0) - \rho_l(0) = \left(\sum_{i=1}^{m} w^u_i   \rho^u_i(0) \right)-\rho_{l}(0)
    \; ,
\end{equation}
where $\rho_{u(l)}(0)$  is the density at the origin of the upper (lower) level, $w^u_i$ is the weight of the configuration $i$ in the upper level CSF-expansion (with $\sum_{i}^m w^u_i \approx 1$),
and $\rho^u_i(0)$, the associated density at the origin. \\
Considering the sixth $J=9/2$-odd parity level, the leading configurations are $5d^76s6p$, $5d^86p$ and $5d^6 6s^26p$. We expect these three configurations to have significantly different electronic densities at the origin, according to their respective occupation number of the $s$ orbitals~\cite{Schetal:2020a}. Thanks to a simple occupation number-based analysis, we expect the following inequalities
\begin{equation}
    \rho(5d^86p)<\rho(5d^76s6p)<\rho(5d^6 6s^26p) 
\end{equation}
to hold, in which the density increases with the occupation number of the $6s$ orbital. Independent Dirac-Fock calculations for each configuration, assuming a common [Xe]$4f^{14}$ core, confirm our intuitions since the computed electron densities, relative to the lowest value, are, in units of $a_0^{-1}$, $\rho(5d^86p)=0 < \rho(5d^76s6p)=103 < \rho(5d^6 6s^26p)=230$. Note that with respect to the $5d^76s6p$ leading configuration, the other two have opposite effects, that can counteract each other as a contamination by the $5d^86p$ configuration decreases the electron density while a contamination by the $5d^66s^26p$ configuration increases it. From a single electron density (or equivalently from a single field shift parameter), it is impossible to recover the relative weight of each configuration, as the determination of the mixing coefficients exhibits an extra degree of freedom (one FS value and two independent mixing coefficients). \\

\subsection{Large-scale calculations}
\label{sec:LSC}
The lack of theoretical calculations for atomic elements with open $d$-shells above the half-shell occupation is due to the fast growing number of CSFs when allowing electron substitutions from the $5d^n$, $n\geq 5$ shell. If single and double substitutions are allowed from $5d^n$, configurations such as $5d^{(n-2)}n_1 l_1 n_2 l_2$ are generated, and that is without considering the outermost electrons. As an example the single configuration built from the substitution $5d^76s6p \to 5d^55f^26s6p$ generates almost 4~000 CSFs, due to the large available number of coupling trees leading to the total angular momentum $J=9/2$. In addition to the intrinsic difficulties related to the open $d$-shells elements, the MCDHF method as implemented in {\sc Grasp} usually requires to generate CSFs by allowing electron substitutions from a group of reference configurations, forming the multi-reference (MR) space, that accounts for most of the static (or nondynamical) correlation \cite{Fischer2016,SinOks:68a}. Considerations presented in Sections~\ref{sec:even} and~\ref{sec:sharing} together with the analyses from Refs.~\cite{Kleef1957,XU200752,Sawatzky:1989aa} lead us to define the following MR: 
\begin{equation}
    \begin{aligned}
    \mathrm{even: }& \ 5d^76s^2, \ 5d^86s, \ 5d^9 \\
    \mathrm{odd: }&\ 5d^76s6p,\ 5d^86p, \ 5d^6 6s^26p
    \end{aligned}
\end{equation}
with the hope of describing
adequately both the lower part of the even spectrum with the close-lying $5d^76s^2$ and $5d^86s$ configurations and the strong configuration mixing among the odd-parity levels. The energy functional~(\ref{eq:FEOL}) was built as a weighted sum over the two lowest even $J=9/2$ levels and the fifteen lowest odd $J=9/2$ levels, to ensure the best possible description of the targeted energy spectrum.

The second step of the calculations, i.e., the generation of the orbital basis set, is performed \textit{layer-by-layer}~\cite{Schetal:2020a}. An individual layer consist of at most one correlation orbital per $l$-symmetry. The sequence of correlation orbital layers are $\mathcal{L}_1=\{5f,5g\}$, $\mathcal{L}_2=\{6d,6f,6g\}$, $\mathcal{L}_3=\{7s,7p,7d,7f,7g\}$, $\mathcal{L}_4=\{8s,8p,8d,8f,8g\}$ and $\mathcal{L}_5=\{9s,9p,9d,9f,9g\}$. When optimising the correlation orbitals belonging to the layer $\mathcal{L}_i$, all orbitals belonging to $\mathcal{L}_j$ $j< i$ and the spectroscopic orbitals are frozen, hence the appellation \textit{layer-by-layer}. For each layer $\mathcal{L}_i$, the corresponding CSFs list was generated by allowing single and double substitutions from the $6s$ and $6p$ orbitals to the active set of orbitals $\{5d,6s,6p\} \bigcup \mathcal{L}_1 \bigcup \dots \bigcup \mathcal{L}_{i}$. The energies of the states of interest are monitored along the increasing number of correlation layers. The total binding energy of the targeted state monotony decreases until the difference between two consecutive correlation layers reaches a few cm$^{-1}$. The convergence of excitations energies, however, as they require a balanced description of the electron correlation between the even and odd parities, do not show necessarily a monotonic convergence but might present oscillations. Nevertheless, the amplitude of the oscillations are decreasing with the number of correlation layers and therefore the calculations can be considered as converged and we may proceed to the next step. RCI calculations are then performed for an enlarged CSFs space generated by allowing SD substitutions from the $5d$ orbital to the largest active set of orbitals, i.e., $\{5d,6s,6p\}  \bigcup_{i=1}^{5}\mathcal{L}_i$. This type of substitutions, although necessary according to the great sensitivity of the level ordering to the $5d$ orbital as already mentioned in Section~\ref{sec:even}, is extremely costly. Note also that the number of CSFs is growing significantly faster for the odd parity (around 1~150~000 CSFs) than for the even parity (around 150~000 CSFs), which might lead to an unbalanced treatment of electron correlation, and therefore a transition energy in poor agreement with experiments. Finally, S substitutions from the $4f$, $5s$ and $5p$ orbitals to the $\{5d,6s,6p\}  \bigcup_{i=1}^{5}\mathcal{L}_i$ active set of orbitals are included, as S substitutions play an important role, due to the one-body structure of the field shift operator~\cite{Filippin2017}. Moreover, they are usually computationally 'cheap' as the number of CSFs generated is small compared to e.g., D substitutions.

\subsection{Root-flipping}
\label{sec:flipping}
We showed in Section~\ref{sec:sharing} that the configuration composition is a key component when computing the electronic field shift factors. This also implies that the $F_{2i,u}$ factors might differ substantially between two upper, odd levels, even though these levels are close in energy. Therefore, we must pay a careful attention that the computed level corresponds to the level involved in the measured transition, i.e., that its absolute position in the theoretical Hamiltonian spectrum of the ordered roots for the considered  $(\Pi,J)$ block symmetry is the same as the experimental energy ordering. The inversion of two energy levels compared to experiments is referred to as \textit{root-flipping}~\cite{Vaeck_1991}. A case of root-flipping appeared in the $^{1}\!F$ series of Ca~II studied by Vaeck \textit{et al.}~\cite{Vaeck_1991},  
and was attributed to the omission of the core polarization that is usually captured by including core-valence correlation effects. This issue was later resolved by improving the correlation model, including core-valence and core-core correlation~\cite{Sundholm1993b}. The influence of root-flipping on isotope shifts was demonstrated by
Aspect \textit{et al.}~\cite{Aspect_1991} through a detailed comparison of theoretical calculations and measurements in Ca~I. The presence of a root-flipping in the theoretical calculations therefore suggests that the correlation model is not optimal and that improvements might restore the correct ordering of the Hamiltonian roots. From a computational point of view, this is far from trivial when the energy spectrum is dense and highly mixed, as it is for Ir~I~\cite{NIST_ASD,moore1958}. In order to support the identification of the targeted level, two means are employed. The first one is based on the Land\'e $g$ factor and the second one on \textit{phase tracking}, i.e., the analysis of the relative phases of the CSFs weights building the ASF.  \\

The Land\'e $g$ factor~\cite{Cowan1981} is an atomic parameter that characterizes the response of the system to an external magnetic field (for more details see Ref.~\cite{Li2020}, and references therein). In $LS$-coupling, the $g$ factor for a state of total angular momentum $J$
\begin{equation}
\label{eq:gLS}
\begin{aligned}
g_{J}&=\sum_{LS}w(LS)g_{J}(LS)\\
&=\sum_{LS}w(LS)\left(1+(g_{s}-1)\frac{J(J+1)+S(S+1)-L(L+1)}{2J(J+1)}\right) \ ,
\end{aligned}
\end{equation}
where $g_s\approx 2$ is the electron $g$ factor~\cite{Cowan1981} and $w(LS)$ are the weights of the $LS$-terms, is a weighted sum over all possible $LS$ terms. Therefore, the $g$ factor only depends on the relativistic mixing and on $L$, $S$ and $J$ angular momenta. Analogously to the sharing rule, it reveals information about the $LS$-composition of the state. The knowledge of the $g$ factors is therefore useful to distinguish two states close in energy, if their $g$ factors are significantly different~\cite{Fischer2001}. In the MCDHF framework of {\sc Grasp},  $jj$-coupling is adopted and the $g$ factors are computed as
\begin{equation}
\label{eq:g}
    g_{\gamma J}=\sum_{i,j}c_i^\gamma c^\gamma_j \frac{1}{2\mu_B}\frac{\langle \gamma_i J||\textbf{N}^{(1)}||\gamma_j J\rangle}{\sqrt{J(J+1)(2J+1)}}
\end{equation}
where $\mu_B$ is the Bohr magneton and $\textbf{N}^{(1)}$ is the relativistic tensorial operator of rank 1 used to describe the interaction between electrons and external magnetic fields~\cite{Cheng1985}. In this work, the fully relativistic expression~(\ref{eq:g}) is used rather than the $LS$-coupling expression~(\ref{eq:gLS}) to estimate $g$ factors, even though the program {\sc jj2lsj} developed by Gaigalas \textit{et al.}~\cite{Gaigalas2003} performs the angular transformation from the $jj$ to the $LSJ$ representation of the ASF that would allow us to use Eq.~(\ref{eq:gLS}). The duality of Eqs.~(\ref{eq:gLS}) and~(\ref{eq:g}) is worthwhile to understand because, although not explicitly visible in its fully relativistic expression, $g$ factors highly depend on the $LS$ state compositions. $g$ factors were measured for Ir~I even and odd states by Van Kleef~\cite{Kleef1957}, that will therefore be used to identify and track the states of interest in this work. \\

Another mean to ensure the correct correspondence between a computed level and a measured level is an analysis of the relative phases of the dominant CSFs in the ASF expansion that we will refer to as phase tracking. The difference between two consecutive RCI calculations performed with the same orbital basis is an enlarged CSFs list. We can decompose the largest of the two CSFs lists as a "small(er)" calculation containing $N_{\text{CSF}}$ CSFs and a "larg(er)" calculation containing the total $N'_{\text{CSF}}$ CSFs, such that $N_{\text{CSF}}<N'_{\text{CSF}}$ and $N'_{\text{CSF}}-N_{\text{CSF}}$ corresponds to the number of newly added CSFs. Going from one calculation to the other, the level energy $E_\gamma$ and the corresponding expansion coefficients $c^\gamma_i  (i=1,\dots,N_{\text{CSF}})$, have changed, due to their interaction with the newly added CSFs, even though the $N_{\text{CSF}}\times N_{\text{CSF}}$ Hamiltonian (sub)matrix  is identical in both calculations. 
If we describe the coefficients $c^\gamma_i$ as $e
^{\mathrm{i}\phi^\gamma_{i}} |c^\gamma_i|$ where $\phi_i=0,\pi$ is the phase associated to the coefficient $c^k_i$, then the relative phase $|\phi^\gamma_{i}-\phi^\gamma_j|$ of two CSFs $i$ and $j$ ($i,j \leq N_{\rm CSF}$) in the eigenvector representing the same level $\gamma$ should remain the same. 
Indeed, the relative phase between two dominant CSFs are usually unaffected when additional CSFs are included. From a pertubative perspective, the addition of CSFs that are energetically less important than the dominant CSFs, e.g., those belonging to the MR, should not modify the relative phases of the leading CSFs. Therefore, if the phase of at least one of the leading CSFs has changed with the introduction of a small perturbation, we interpret this as the signature of a root-flipping. This reasoning is illustrated in~\ref{app:rootflipping} using simple matrices to mimic the behaviour observed in the large-scale RCI calculations. 

The observation $|\phi^\gamma_{i}-\phi^\gamma_{j}|\neq |\phi'^{\gamma} _{i}-\phi'^{\gamma} _{j}|$, ($i,j \leq N_{\rm CSF}$) is therefore a strong sign of a root-flipping. In the simplest case with two configurations, each generating one CSF, the $2\times 2$ Hamiltonian matrix has two eigenvalues $E_1$ and $E_2$, associated to two normalized eigenvectors with coefficients $(c^1_1,c^1_2)$ and 
$(c^2_1,c^2_2)=(c^1_2,-c^1_1)$, respectively. The relative phases of the first and second eigenvectors between their two components are therefore $|\phi^1_{1}-\phi^1_2|=0$ and $|\phi^2_{1}-\phi^2_2|=1$, respectively.

\section{Results and discussions}
\label{sec:results}
\subsection{Field shifts factors}
\label{sec:FSfac}
In this section, we present the results of our largest calculations, i.e., MR calculations from which SD substitutions are allowed from the $5d$, $6s$ and $6p$ orbitals and S substitutions are allowed from the $4f$, $5s$ and $5p$ orbitals. The twelve lowest odd $J=9/2$ states are investigated to demonstrate the complexity of the iridium spectrum. The even part of the spectrum is less interesting, although it is worth mentioning that the excitation energy of the first excited even $J=9/2$ state is 5~728~cm$^{-1}$, which is almost 3~000 cm$^{-1}$ larger than the experimental value. Nevertheless, we verified that the ground state was indeed the $5d^7 6s^2 \ ^{4}\!F_{9/2}$, in contrast with the less elaborate calculations described in Section~\ref{sec:even}. As discussed in the latter, the low even parity part of the spectrum was demonstrated to be highly sensitive to the correlation model. The introduction of the D substitutions from the $5d$ shell in the present large-scale calculations is necessary to obtain the correct, experimentally measured, level ordering of  the $5d^7 6s^2 \ ^{4}\!F_{9/2}$ and $5d^8 6s \ ^{4}\!F_{9/2}$ levels. A correct description of the ground state is crucial since the $5d^7 6s^2 \ ^{4}\!F_{9/2} \rightleftharpoons 5d^8 6s \ ^{4}\!F_{9/2}$ ground state inversion would have a dramatic effect on the electronic field shift parameters according to the sharing rule presented in Section~\ref{sec:sharing}.

Table~\ref{tab:MR} shows $E_\gamma$, the excitation energy of the $\gamma$th level ($\gamma=1,\dots,12$) in the $J^\Pi=(9/2)^{\rm o}$ symmetry block, and $\Delta E_\gamma$, its discrepancy with observation, in columns 2 and 3, respectively. Since the energy of the even states, including the ground state, are not highly accurate and since we expect an unbalanced description of the even-odd spectrum, we also provide $E^{\rm o}_\gamma$, the excitation energy of the $\gamma$th ($\gamma=1,\dots,12$) odd state relative to the lowest odd state, as well as $\Delta E^{\rm o}_\gamma$, its discrepancy with observation, in columns 4 and 5, respectively. One  observes that $|\Delta E^{\rm o}_\gamma| < |\Delta E_\gamma|$ for all levels, but the sixth one. The reasonably good agreement with experiments of the odd-parity spectrum but the overall poor agreement for the excitations energies confirms an unbalanced treatment of the electron correlation between the two parities.
\begin{table}[!h]
    \centering
        \caption{The energy of the $J=9/2$ odd levels, which are identified by their key number $\gamma$, are given relative to the ground state ($E_\gamma$) and to the lowest odd state ($E_\gamma^{\rm o}$), as well as their respective differences with experiments ($\Delta E_\gamma$ and $\Delta E_\gamma^{\rm o}$, respectively). For each odd level $\gamma$, the weights of the leading configurations are given, together with the associated $g$ factor and its deviation from the experimental value, which is taken from Refs.~\cite{Kleef1957,NIST_ASD}. Finally, the electronic field shift factor, $\Delta F_{0,k}$, is given for each transition $k$ between the ground state and the $\gamma$th $J=9/2$ odd level. According to Eq.~(\ref{eq:F2i}), a negative value for $\Delta F_{0,k}$ indicates a loss of electronic density associated with the absorption 
 $5d^{7}6s^{2} \ ^{4}\!F_{9/2} \rightarrow \gamma (9/2)^o$.
}
    \label{tab:MR}
    \begin{tabular}{ccccccccccccc}
     \hline
    \hline
    & \multicolumn{4}{c}{Energies (cm$^{-1}$)} && \multicolumn{3}{c}{Composition ($\%$)}& & &(GHz/fm$^2$)\\
    \cline{2-5}
    \cline{7-9}
    \cline{12-12}
    $\gamma$ & $E_\gamma$ & $\Delta E_\gamma$ & $E_\gamma^{\rm o}$ &$\Delta E_\gamma^{\rm o}$&& $5d^7 6s6p$ & $5d^86p$ & $5d^6 6s^26p$ & $g$ & $\Delta g$ &$\Delta F_{0,k}$  \\
    \hline
1 &      22220& $-$4087 &      0      &      0&   &    89.6& &          2.9 &  1.48&    0.00 &   $-$31.03\\
2 &      28357& $-$4156 &      6137&         $-$69 &   &    87.0    &  &     5.7 &  1.37  &  0.01  &  $-$30.10\\
3 &      31713& $-$3368 &      9493  &        720& &  68.7 &  7.3 &  15.9 &  1.27  &  0.00 &   $-$28.29   \\
4 &      34563& $-$3309 &      12342&       778 && 76.8  & 10.0  & 5.1  & 1.27  &  0.02   & $-$36.16   \\
5 &      36187& $-$2043 &      13967&        2044&& 56.5 &  26.3 &  9.3 &  1.27  &  0.05  &  $-$41.19   \\
6 &      37281& $-$3109 &      15060&        978 && 61.9 &  0.3 &  29.4 &  1.38  &  0.15  &  $-$17.05  \\
7 &      38124& $-$2995 &      15903&        1092& &35.8 &  36.0 &  19.7 &  1.22  &  $-$0.15& $-$42.20  \\
8 &      38739& $-$3541 &      16518&        546 && 42.6 &  27.8 &  20.9 &  1.29  &  $-$0.09& $-$36.41 \\
9 &      41399& $-$3253 &      19179&        834 && 74.9 &  4.4 &  12.7 &  1.23  &  $-$0.02& $-$29.14 \\
10&      42633& $-$3587 &      20413&        500 && 67.3 &  4.2 &  19.9 &  1.34  &  0.11    &$-$23.64 \\
11&      43364& $-$3008 &      21143&        1079 && 71.7 &  10.7 &  9.3 &  1.27  &  $-$0.09 &$-$35.77\\
12&      43962& $-$3243 &       21742&        844 && 77.6 &  11.1 &  3.3 &  1.06  &  0.01    &$-$29.55 \\
    \hline
    \hline
    \end{tabular}
\end{table}

Table~\ref{tab:MR} also displays, for each odd level, the configuration compositions (in~$\%$) of the three leading configurations i.e., $5d^7 6s6p$, $5d^86p$ and $5d^6 6s^26p$. Note that the sum of all contributions is not 100$\%$, as only the leading contributions are taken into account. The contributions of the $5d^86p$ and $5d^6 6s^26p$ configuration reach $36.0\%$ and $29.4\%$ for the sixth and seventh levels, respectively. The lowest odd-parity state, on the other hand is almost unaffected by the configuration mixing and is highly dominated by the $5d^76s6p$ configuration. From the Dirac-Fock (DF) densities computed in Section~\ref{sec:sharing}, we evaluate the following $F_{0}$ factors in GHz/fm$^{-2}$: $F_{0}(5d^86p)=0 < F_{0}(5d^76s6p)=39 < F_{0}(5d^6 6s^26p)=87$. Interestingly, when the three DF $F_{0}$ factors are computed using the same orbital basis, i.e., the 'best' orbital basis, we have, in GHz/fm$^{-2}$, $F_{0}(5d^86p)=0 < F_{0}(5d^76s6p)=39.6 < F_{0}(5d^6 6s^26p)=79.2$. The $F_0$ factor relative to the $5d^86p$ configuration is exactly twice larger for the $5d^6 6s^26p$ configuration than for the $5d^76s6p$ configuration, in agreement with the $6s$ orbital occupation number. Constraining the $5d$, $6s$ and $6p$ orbitals to be the same for the three configurations has a larger impact on the $5d^6 6s^26p$ configuration.  It is therefore consistent to have smaller $\Delta F_{0,k}$ parameters (in absolute value) for transitions including upper levels strongly contaminated by the $5d^6 6s^26p$ configuration such as the 6th level and larger (in absolute value) $\Delta F_{0,k}$ parameters for levels strongly contaminated by the $5d^86p$ configuration such as the 7th level. \\
In this work, we are interested in the field shift parameters of the $5d^{7}6s^{2} \ ^{4}\!F_{9/2}$
$\to (\mbox{odd},J= 9/2)$ line at 247.587~nm. Since the computed transition energies are not fully reliable and since large variations are observed for the FS parameters along the odd levels, the identification of the levels involved in the transition is performed carefully. The ground, even state, as stated earlier, is well-defined and well identified. On the contrary, the upper, odd, level involved in the transition is not. The ordered roots of the  Hamiltonian would suggest the matching between the 6th theoretical root and the 6th observed level, corresponding to the targeted level in~\cite{Muketal:2020a}. However, looking at the $g$ factors displayed in Table~\ref{tab:MR} and their difference with the experimental values from Ref.~\cite{Kleef1957}, the 6th and 7th levels disagree with experiments by $+$0.15 ($12\%)$ and $-$0.15 ($11\%$), respectively. Considering a root-flipping between the 6th and 7th levels would substantially reduce theory-observation discrepancy  to $-$0.01 ($0.7\%$) and $+$0.01 ($0.6\%)$, respectively. Taking this argument into account, the $\Delta F_{0,247}$ parameter becomes $-42.20$~GHz/fm$^2$ instead of $-17.05$~GHz/fm$^2$.

To further confirm the root-flipping, one can look at the phases of the leading CSFs in two sets of calculations. In Table~\ref{tab:c_i}, we present the mixing coefficients of the leading CSFs ($|c_i|>0.1$) in $LS$-coupling for two different RCI calculations. Both include SD substitutions from the $5d$, $6s$ and $6p$ orbitals but only the second one includes S substitutions from the $4f$, $5s$ and $5p$ orbitals. Each CSF is given with its coupling tree to ensure its uniqueness along with its corresponding mixing coefficients in the 6th and 7th eigenvectors of the Hamiltonian, respectively noted $c^{6}$ and $c^{7}$. Comparing $c^6$ in both calculations, one obviously observes that (i) some CSFs are missing in the second calculations, (ii) even when introducing a global phase $-\pi$, the fourth CSF has a different phase in the two calculations and (iii) the $^6\!D$ character is strong in the second calculation while absent in the first calculation. Comparing $c^6$ of the first calculation with $c^7$ from the second one, most of the dominant CSFs are found in both calculations, with the correct phases. Some of them are missing due to the cutoff ($|c_i|>0.1$). Choosing the leading CSFs of 6th eigenvector of the smaller calculation (i.e., before root-flipping) as reference, Table~\ref{tab:phi} displays the corresponding phases of these CSFs in the 6th and 7th levels computed with the S from the $4f$, $5s$ and $5p$ orbitals. This small and simple analysis demonstrates that when including the single substitutions the 6th and 7th levels indeed exchange their position in the theoretical Hamiltonian roots ordering. Note that the same analysis could be performed in $jj$-coupling.

\begin{table}[!h]
    \centering
    \caption{Mixing coefficients of the leading CSFs for the 6th ($c^{6}$) level for two different RCI calculations. Both include SD substitutions from the $5d$ orbital but only the second one includes S substitutions from the $4f$, $5s$ and $5p$ orbitals. In the latter, the composition of the 7th level is also shown. Only the CSFs with $|c_i| > 0.1$ are displayed.}
    \label{tab:c_i}
    \hspace*{-1cm}
    \begin{tabular}{ccrc|ccrcr}
    \hline
    \hline
            &\multicolumn{2}{c}{SD$(5d)$} & & & \multicolumn{4}{c}{ SD$(5d)$ + S$(4f,5s,5p)$} \\
            \cline{2-3}
            \cline{6-9}
        &CSF$^{(6)}$ & \multicolumn{1}{c}{$c^{6}$} &&& CSF$^{(6)}$&\multicolumn{1}{c}{$c^{6}$} & CSF$^{(7)}$& \multicolumn{1}{c}{$c^{7}$}  \\
        \hline 
1 &$5d^8(^{3}\!F_2)6p \ ^{2}\!G       $     & 0.4888    &&& $5d^6(^{5}\!D_4)6s^26p \ ^{6}\!D $        &0.4429    &$5d^8(^{3}\!F_2)6p \ ^{2}\!G       $     &0.4686\\
2 &$5d^7(^{4}\!F_3)6s(^{5}\!F)6p \ ^{6}\!G $& 0.4125    &&& $5d^7(^{4}\!P_3)6s(^{5}\!P)6p \ ^{6}\!D $ &$-$0.4202 &$5d^7(^{4}\!F_3)6s(^{3}\!F)6p \ ^{4}\!F $&0.4155\\
3 &$5d^7(^{4}\!F_3)6s(^{3}\!F)6p \ ^{4}\!F $& 0.2764    &&& $5d^7(^{4}\!F_3)6s(^{5}\!F)6p \ ^{6}\!G $ &$-$0.2793 &$5d^6(^{5}\!D_4)6s^26p \ ^{4}\!F $       &0.3388\\
4 &$5d^7(^{4}\!F_3)6s(^{3}\!F)6p \ ^{2}\!G $& $-$0.2626 &&& $5d^7(^{4}\!F_3)6s(^{5}\!F)6p \ ^{4}\!F $ &0.2564    &$5d^8(^{3}\!F_2)6p \ ^{4}\!F       $     &0.2580\\
5 &$5d^8(^{3}\!F_2)6p \ ^{4}\!G       $     & 0.2307    &&& $5d^7(^{4}\!F_3)6s(^{3}\!F)6p \ ^{2}\!G $ &0.2296    &$5d^8(^{3}\!F_2)6p \ ^{4}\!G       $     &0.2393\\
6 &$5d^8(^{3}\!F_2)6p \ ^{4}\!F       $     & 0.2228    &&& $5d^6(^{5}\!D_4)6s^26p \ ^{6}\!F $        &$-$0.2054 &$5d^7(^{2}\!G_3)6s(^{1}\!G)6p \ ^{2}\!G $&0.2006\\
7 &$5d^7(^{4}\!F_3)6s(^{3}\!F)6p \ ^{4}\!G $& $-$0.2165 &&& $5d^7(^{4}\!F_3)6s(^{5}\!F)6p \ ^{4}\!G $ &0.1947    &$5d^6(^{3}\!F_2)6s^26p \ ^{2}\!G $       &$-$0.1495\\
8 &$5d^7(^{2}\!G_3)6s(^{1}\!G)6p \ ^{2}\!H $& $-$0.1638 &&& $5d^7(^{4}\!F_3)6s(^{5}\!F)6p \ ^{6}\!D $ &0.1569    &$5d^7(^{2}\!G_3)6s(^{3}\!G)6p \ ^{4}\!G $&0.1464\\
9 &$5d^7(^{2}\!G_3)6s(^{1}\!G)6p \ ^{2}\!G $& 0.1451    &&& $5d^7(^{4}\!F_3)6s(^{3}\!F)6p \ ^{4}\!G $ &0.1544    &$5d^7(^{4}\!F_3)6s(^{5}\!F)6p \ ^{6}\!F $&$-$0.1428\\
10&$5d^6(^{5}\!D_4)6s^26p \ ^{4}\!F $       & 0.1353    &&& $5d^7(^{2}\!G_3)6s(^{3}\!G)6p \ ^{4}\!F $ &$-$0.1488 &$5d^7(^{2}\!G_3)6s(^{3}\!G)6p \ ^{4}\!F $&$-$0.1269\\
11&$5d^7(^{4}\!F_3)6s(^{5}\!F)6p \ ^{4}\!G $& $-$0.1169 &&& $5d^7(^{4}\!F_3)6s(^{3}\!F)6p \ ^{4}\!F $ &0.1457    &$5d^6(^{3}\!F_4)6s^26p \ ^{4}\!F $       &$-$0.1240\\
12&$5d^7(^{2}\!G_3)6s(^{3}\!G)6p \ ^{2}\!G $& $-$0.1168 &&& $5d^6(^{3}\!F_4)6s^26p \ ^{4}\!F $        &$-$0.1426 &$5d^7(^{4}\!F_3)6s(^{3}\!F)6p \ ^{4}\!G $&$-$0.1194\\
13&                                         &           &&& $5d^7(^{2}\!G_3)6s(^{1}\!G)6p \ ^{2}\!H $ &0.1398    &$5d^7(^{2}\!H_3)6s(^{1}\!H)6p \ ^{2}\!G $&0.1098\\
14&                                         &           &&& $5d^7(^{2}\!D_3)6s(^{3}\!D)6p \ ^{4}\!F $ &0.1312    &$5d^7(^{4}\!F_3)6s(^{5}\!F)6p \ ^{6}\!G $&0.1054\\
15&                                         &           &&& $5d^7(^{2}\!D_1)6s(^{3}\!D)6p \ ^{4}\!F $ &0.1306    &$5d^8(^{1}\!G_2)6p \ ^{2}\!G       $     &0.1023\\
16&                                         &           &&& $5d^7(^{2}\!G_3)6s(^{3}\!G)6p \ ^{4}\!G $ &0.1229    &$5d^6(^{3}\!F_4)6s^26p \ ^{2}\!G $       &$-$0.1013\\
17&                                         &           &&& $5d^6(^{5}\!D_4)6s^26p \ ^{4}\!F $        &$-$0.1151 &\\
         \hline
         \hline
    \end{tabular}
\end{table}

\begin{table}[!h]
    \centering
    \caption{Phases of the leading CSFs for the 6th ($\phi^{6}$) and 7th ($\phi^{7}$) levels for two different RCI calculations. Both include SD substitutions from the $5d$ orbital but only the second one includes S substitutions from the $4f$, $5s$ and $5p$ orbitals. For few CSFs, no phase is displayed as their expansion coefficient was too small to be taken into account into the $jj$ to $LSJ$ transformation.}
    \label{tab:phi}
    \begin{tabular}{ccccc}
    \hline
    \hline
            & SD$(5d)$ & & \multicolumn{2}{c}{ + S$(4f,5s,5p)$} \\
            \cline{2-2}
            \cline{4-5}
        CSF & $\phi^{6}$ && $\phi^{6}$ & $\phi^{7}$  \\
        \hline 
$5d^8(^{3}\!F_2)6p \ ^{2}\!G       $& 0 && / & 0 \\
$5d^7(^{4}\!F_3)6s(^{5}\!F)6p \ ^{6}\!G $& 0 && $\pi$ & 0 \\
$5d^7(^{4}\!F_3)6s(^{3}\!F)6p \ ^{4}\!F $& 0 && 0 & 0 \\
$5d^7(^{4}\!F_3)6s(^{3}\!F)6p \ ^{2}\!G $& $\pi$ && 0 & $\pi$ \\
$5d^8(^{3}\!F_2)6p \ ^{4}\!G       $& 0 && / & 0 \\
$5d^8(^{3}\!F_2)6p \ ^{4}\!F       $& 0 && / & 0 \\
$5d^7(^{4}\!F_3)6s(^{3}\!F)6p \ ^{4}\!G $& $\pi$ && 0 & $\pi$ \\
$5d^7(^{2}\!G_3)6s(^{1}\!G)6p \ ^{2}\!H $& $\pi$ && 0 & $\pi$ \\
$5d^7(^{2}\!G_3)6s(^{1}\!G)6p \ ^{2}\!G $& 0 && / & 0 \\
$5d^6(^{5}\!D_4)6s^26p \ ^{4}\!F $& 0 && $\pi$ & 0 \\
         \hline
         \hline
    \end{tabular}
\end{table}

\subsection{Sensitivity to the Breit and QED interactions}
\label{sec:BQ}
The configuration compositions presented in Table~\ref{tab:MR} showed that the levels 5, 6, 7 and 8 are highly mixed. They are also close in energy, as these four levels belong to an energy interval of $2~550$ cm$^{-1}$. We can therefore expect strong interactions between these states. We investigate the sensitivity of the configuration mixing along these states by comparing RCI calculations that included the Breit interaction and QED corrections and calculations that neglect them. These effects are known to be large for heavy elements~\cite{Indelicato:2007aa}, and can affect strongly atomic properties~\cite{Li2012c}. As already discussed in Section~\ref{sec:sharing}, since the configuration mixing is directly reflected in the field shift parameters, we will rather compare field shift parameters, $\Delta F^{\rm BQ}_{0,k}$, estimated with the Breit interaction and QED corrections in the Hamiltonian, and $\Delta F_{0,k}$ omitting these corrections. In Table~\ref{tab:BQ} are shown $\Delta F^{\rm BQ}_{0,k}$ and $\Delta F_{0,k}$ in GHz/fm$^{-2}$, as well as their difference $\delta \Delta F^{\rm BQ}_{0,k}= \Delta F^{\rm BQ}_{0,k} - \Delta F_{0,k}$, for each transition $k$ between the ground state and the $\gamma$th odd-parity $J=9/2$ level.
\begin{table}[!h]
    \centering
    \caption{Electronic field shift parameters $\Delta F_{0,k}$ and $\Delta F^{\rm BQ}_{0,k}$ computed without and with Breit interaction and QED corrections, respectively, and their difference $\delta\Delta F^{\rm BQ}_{0,k}$, in GHz/fm$^2$. They are given for each transition $k$, which is the transition between the ground state and the $\gamma$th $J^\Pi=(9/2)^{\rm o}$ level.}
    \label{tab:BQ}
    \begin{tabular}{cccc}
    \hline
    \hline
     $\gamma$  & $\Delta F_{0,k}$ & $\Delta F^{\rm BQ}_{0,k}$ & $\delta\Delta F^{\rm BQ}_{0,k}$\\
    \hline
1	& $-$31.03	& $-$31.15	& $-$0.11 \\
2	& $-$30.10	& $-$30.43	& $-$0.33 \\
3	& $-$28.29	& $-$29.69	& $-$1.40 \\
4	& $-$36.16	& $-$36.89	& $-$0.73 \\
5	& $-$41.19	& $-$44.91	& $-$3.72 \\
6	& $-$17.05	& $-$19.65	& $-$2.60 \\
7	& $-$42.20	& $-$48.08	& $-$5.87 \\
8	& $-$36.41	& $-$28.21	& 8.20 \\
9	& $-$29.14	& $-$29.01	& 0.12 \\
10	& $-$23.64	& $-$22.51	& 1.13 \\
11	& $-$35.77	& $-$35.73	& 0.05 \\
12	& $-$29.55	& $-$29.90	& $-$0.35 \\
    \hline
    \hline
    \end{tabular}
\end{table}
 The results are surprising as the larger the configuration mixing is, the more sensitive is the corresponding field shift parameter. The discrepancy reaches up to $22\%$ for the 8th level and $7\%$ for the targeted 7th (6th considering the root-flipping) level. The lower levels, on the other hand, are almost unaffected, which corresponds to stable configuration compositions. The strong impact of the Breit interaction and QED corrections found in the present work should incite further work in the direction of a full variational treatment of these corrections in the MCDHF orbital optimization process, as it is done in the mcdfgme package~\cite{Indelicato2019}, an alternative to the {\sc Grasp} package.

\subsection{Using the line at 351~nm}
Another experimental paper dedicated to isotope shifts in neutral iridium was published by Verney \textit{et al.,}~\cite{Verney:2006aa}. They reported IS values for a long chain of $^A$Ir isotopes $A=193, 191, 189, 187,186m, 186g, 185, 184, 183, 182$, for the $5d^{7}6s^{2} \ ^{4}\!F_{9/2} \to 5d^{7}6s6p \ ^{6}\!F^{\rm o}_{11/2}$ transition at 351.5~nm, where the $^{6}\!F^{\rm o}_{11/2}$ level is the lowest of the $J^\Pi=(11/2)^{\rm o}$ symmetry. They reported a field shift parameter of $\Delta F_{0,351}= -31.94$~GHz/fm$^{2}$, computed using the relativistic multiconfiguration Dirac-Fock code of Desclaux~\cite{DESCLAUX1975}. No details of the calculations such as the correlation models or the orbital optimization strategy were reported, although the given Refs.~\cite{Ulm:1986aa,WALLMEROTH1989,Hilberath:1992aa} suggest that the present calculations are more elaborate. Using the expression~(\ref{eq:F0}) for the field shifts of both transitions, we have
\begin{equation}
     \frac{\delta \nu^{AA'}_{{\rm FS},351}}{\Delta F_{0,351}}= \frac{\delta \nu^{AA'}_{{\rm FS},247}}{\Delta F_{0,247}} \ .
\end{equation}
Using the (193,191) isotope pair for which the isotope shifts were measured for both lines at $\lambda= 247$ and 351~nm and considering $\delta \nu_{{\rm MS},k} \ll  \delta \nu_{{\rm FS},k}$, one can estimate $\Delta F_{0,247}=-49.15$~GHz/fm$^2$, which is close to $\Delta F^{\rm BQ}_{0,247}=-48.08$~GHz/fm$^2$ presented in Table~\ref{tab:BQ}. Note that we used expression~(\ref{eq:F0}) and not~(\ref{eq:F2i}) for the field shift, which is consistent with the approach adopted in Ref.~\cite{Verney:2006aa} where the isotope shifts are calculated according to
\begin{equation}
    \label{eq:Flambda}
    \delta\nu_{\text{FS},k}^{AA'}= \frac{\Delta F_{0,k}}{h}\lambda^{AA'} \ ,
\end{equation}
with $\lambda^{AA'}$ being a nuclear parameter that includes higher order nuclear moments. The good agreement between the two methods of calculations might be surprising at first, considering that the field shift factor of the line at 351~nm was obtained within a less sophisticated computational model. Nevertheless, assuming that our observations along the $J^\Pi=(9/2)^{\rm o}$ spectrum also apply to the $J^\Pi=(11/2)^{\rm o}$ spectrum, then, as we demonstrated in Sections.~\ref{sec:FSfac} and~\ref{sec:BQ}, the lowest state of a given parity is almost unaffected by both the configuration mixing and the Breit interaction and QED corrections. Since the transition at 351~nm involves the lowest $J^\Pi=(11/2)^{\rm o}$ level, the need for an adequate treatment of the above effects is most likely less critical.
\subsection{Final values}
The present analysis, based on the sharing rule, the acceptation of a root-flipping, the inclusion of the Breit interaction and QED corrections, also supported by the observed field shift of the line at 351~nm, leads us to suggest the following values for the field shift parameters
\begin{equation}
    \label{eq:res}
    \begin{aligned}
    \Delta F_{0,247} &= -48.1(3.0)~\text{GHz/fm}^2 \ ,\\
    \Delta F_{2,247} &=0.0526~\text{GHz/fm}^4 \ ,   \\
    \Delta F_{4,247} &=-0.000150~\text{GHz/fm}^6 \ ,  \\
    \Delta F_{6,247} &=0.000000267~\text{GHz/fm}^8 \ . \\
    \end{aligned}
\end{equation}
As detailed in~\cite{Muketal:2020a}, Eq.~(\ref{eq:F2i}) was used to expand the observed field shifts of the  247~nm spectra line in products of the present electronic factors (\ref{eq:res}) and  differences in radial moments of the nuclear 
charge distribution, to determine the axially symmetric quadrupole deformation $\beta^A_2 $ and $\langle r^2_c \rangle ^A$ values.

The electronic parameters of the reformulated FS of Eq.~(\ref{eq:Fved}) are 
\begin{equation}
\label{eq:resved}
    \begin{aligned}
    \Delta F^{(0),ved}_{0,247} &= -44.9(3.0)~\text{GHz/fm}^2 \ ,\\
    \Delta F^{(1),ved}_{0,247} &=0.00444~\text{GHz/fm}^4 \ ,   \\
    \end{aligned}
\end{equation}
where the corrected $\Delta F^{(0),ved}_{0,247}$ is $7\%$ lower than $\Delta F_{0,247}$.  In this approach,
\begin{equation}
    \label{eq:refapprox}
    \delta\nu_{\text{FS},k}^{AA'} \approx \frac{\Delta F^{(0),ved}_{0,k} }{h} \ \delta\langle r^2 \rangle^{AA'} 
\end{equation}
is by far the leading contribution of Eq.~(\ref{eq:Fved}), which therefore allows us to infer that the effects of a varying electron density inside the nucleus on the isotope shift is on the order of $7\%$, strengthening the need to incorporate this effect. The rounded uncertainty of 3.0~GHz/fm$^2$ is estimated for the $\Delta F_{0,k}$ dominant electronic factor by considering the effect of Breit interaction and QED corrections and the value extracted from the line at 351~nm. The higher-order factors highly depends on each other, and therefore, no reliable and meaningful uncertainties could be estimated. 


\subsection{Variational approach}
The evaluation of the field shift parameters from Eqs~(\ref{eq:F0}),~(\ref{eq:F2i}),~(\ref{eq:ved}), and~(\ref{eq:Fved}) is based on a perturbative approach, in which the ASFs of interest are computed for only one reference isotope. Taking the isotope pair (191,193) as an example, the corresponding resulting FS is $\delta \nu^{193,191}_{\text{FS},247}=-8.1$~GHz\footnote{Adopting the convention of Eq.~(\ref{eq:deltanu}), a negative sign for $\delta \nu^{193,191}$ (or a positive sign for $\delta \nu^{191,193}$) indicates a smaller frequency for the  heavier isotope.} when evaluated using Eq.~(\ref{eq:refapprox}), that combines the mean square radii $\langle r^2 \rangle^{A}=\left( 0.836A^{1/3} + 0.570\right)^2$ with the field shift parameter $\Delta F^{(0),ved}_{0,k}$ from Eq.~(\ref{eq:resved}) estimated using the RCI eigenvectors of the DCB+QED Hamiltonian~(\ref{eq:DCB+QED}).
Omitting the long wavelength Breit interaction and QED corrections reduces the field shift value to $\delta \nu^{193,191}_{\text{FS},247} =-7.1$~GHz.  A 'variational' approach, in which the large-scale calculations are performed for two different isotopes, can be used as an alternative and a validation method. The computational cost, however, is large, as the calculations are performed twice. The isotope field shift is then computed directly according to Eq.~(\ref{eq:deltanu}). The variational FS values are $\delta \nu^{193,191}_{\text{FS},247}$$=-7.7$~GHz and $\delta \nu^{193,191}_{\text{FS},247}$$=-6.7$~GHz with and without the Breit interaction and QED corrections,
respectively, which are $0.4$~GHz smaller than the ones computed with the electronic factors. This confirms the need to include the Breit interaction and QED corrections in our calculations. One could also scale the electronic factors in order to have a perfect agreement with the 'variational' approach. This leads to e.g., $\Delta F_{0,247}=-45.8$~GHz/fm$^2$, which is almost $5\%$ lower than the value presented in the previous section.

\section{Conclusion}

We presented the details of the large-scale \textit{ab initio} atomic calculations of the electronic isotope shift parameters for the $5d^{7}6s^{2} \ ^{4}\!F_{9/2} \to (\mbox{odd},J= 9/2)$ transition line at 247.587~nm, that were recently used to extract nuclear mean square radii and nuclear deformations for heavy isotopes of neutral Ir~\cite{Muketal:2020a}. Large configuration mixing was observed for odd parity levels, which strongly affects  the field shift factors according to the sharing rule. A careful investigations of the Land\'e $g$ factors and relative phases of the leading CSFs unambiguously demonstrated the existence of a root-flipping between the 6th and 7th roots of the $(9/2)^{\rm o}$ Hamiltonian block matrix. The introduction of the Breit interaction and QED interactions illustrated the sensitivity of the highly mixed levels compared to the lower, purer levels. A comparison with the electronic field shift factor obtained for the transition line at 351~nm confirmed the need to include these higher-order effects in the Hamiltonian.

\section*{Acknowledgements}
The authors are grateful to J\"orgen Ekman (Malm\"o University) for enlightening discussions. SS is a FRIA grantee of the F.R.S-FNRS. MG acknowledges support from the FWO and FNRS Excellence of Science Programme (EOSO022818F). 
\clearpage 
\appendix
\section{Root-flipping}
\label{app:rootflipping}
Let us start  from the Hamiltonian matrix
\begin{equation}
{\bf H}^n_{ij} = \langle \Phi_i \vert {\cal H} \vert \Phi_j \rangle \; ,
\end{equation}
evaluated in a given basis of dimension $n$
\begin{equation}
\label{App:basis}
\{  \Phi_1 , \Phi_2, \ldots, \Phi_n \}  \; .
\end{equation}
The addition of a new function $\Phi_{n+1}$ to the set (\ref{App:basis}) increases the dimension of the Hamiltonian matrix of one unit. The comparison of the order of the eigenvalues $\{ \lambda^{n+1}_k \}$ of the matrix ${\bf H}^{n+1} $
\begin{equation}
\label{App:eigen}
{\bf H}^{n+1} {\bf c}^k = \lambda^{n+1}_k {\bf c}^k
\end{equation}
with the 
eigenvalues  $\{ \lambda^{n}_k \}$ of ${\bf H}^{n} $  is the foundation of the well-known Hylleraas-Undheim-McDonald (HUM) interleaving theorem~\cite{HylUnd:30a,McD:33a}.  
 Relative phases of the major components of the eigenvector ${\bf c}^k$ determine the expectation value
$\langle {\cal O} \rangle_k$ associated with the relevant observable as follows
\begin{equation}
\label{App:obs}
\langle {\cal O} \rangle_k = \sum_i \sum_j
c^{k \ast}_i c^k_j \langle \Phi_i \vert {\cal O} \vert \Phi_j \rangle
= \sum_i \sum_j
c^{k \ast}_i c^k_j {\bf O}_{ij} \; .
\end{equation}
Monitoring these relative phases  when increasing the space might be rewarding to  correlate the physical states when extending the basis. \\

Let us compare the eigenvectors and eigenvalues of the two following matrices
\[
{\bf H}^4 = 
\left(
\begin{array}{cccc}
{\bf 1.0} & 0.5 & 0.3 & 0.2 \\
0.5 & {\bf 2.0} & 0.1 & 0.2 \\
0.3 & 0.1 & {\bf 3.0} & 0.2 \\
0.2& 0.2 & 0.2 & {\bf 3.2}
\end{array}
\right)
\]
and 
\[
{\bf H}^6 = 
\left(
\begin{array}{rrrrrrrr}
{\bf 1.0} & 0.5 & 0.3 & 0.2 & \vert & 0.0 & 0.0\\
0.5 & {\bf 2.0} & 0.1 & 0.2 & \vert &  0.0 & 0.0  \\
0.3 & 0.1 & {\bf 3.0} & 0.2 & \vert &  -1.0 & -0.5 \\
0.2& 0.2 & 0.2 & {\bf 3.2} & \vert & -1.0 & -1.0 \\
\hdashline 
0.0 & 0.0 & -1.0 & -1.0 & \vert & {\bf 6.0} & 0.5\\
0.0 & 0.0 & -0.5 & -1.0 & \vert & 0.5 &  {\bf 7.0} \\
\end{array}
\right)
\]
where the matrix ${\bf H}^6$ results from the extension of ${\bf H}^4$ by considering the interaction with two levels relatively isolated from the first four but largely  interacting with the third and fourth levels of the first set. 
In the context of the present paper, we would like to emphasize the root-flipping of the 3rd and 4th eigenpairs due to the ``perturbation'' of ${\bf H}^4$ by the addition of $\Phi_5$ and $\Phi_6$ in the basis  spanning ${\bf H}^6$.
The interleave Hylleraas-Undheim-MacDonald (HUM) theorem \cite{HylUnd:30a,McD:33a} is illustrated by the sequence of the $n=4, 5$ and $6$ eigenvalues $\{ \lambda^n_k \}$  of ${\bf H}^4$, ${\bf H}^5$ and ${\bf H}^6$. 
\[
\begin{array}{ccc}
\lambda^4_k & \lambda^5_k & \lambda^6_k \\
\hline
- & - & 7.7997632 \\
- & 6.6110373 & - \\
{\bf 3.4320493} & - & 5.9840097 \\
- & 2.9742613  & - \\
{\bf 2.8869591} & - & {\bf 2.8813157} \\
- & 2.7904410 & - \\
2.1174171 & - & {\bf 2.7406404} \\
- & 2.0652637 & - \\
0.7635736 & - & 2.0367877 \\
- & 0.7589959 & - \\
- & - & 0.7574886 
\end{array}
\]

\vspace*{1cm}

The spectrum of ${\bf H}^4$ is strongly perturbed by the addition of the two extra basis functions ($\Phi_5$ and $\Phi_6$). 
The root-flipping can be observed by looking at the reorganisation of the 3rd and 4th eigenvectors of ${\bf H}^4$ and ${\bf H}^6$, respectively, that read as the following columns
\[
\begin{array}{rrrr} 
\lambda_3^4 = 2.8869591 & \lambda_4^4 = 3.4320493 & 
\lambda_3^6 = 2.7406404 & \lambda_4^6 = 2.8813157 \\
\hline 
0.070606 & 0.175189 & 0.278179 & 0.030818 \\
0.003494 & 0.210066 & 0.387670 & 0.135137 \\
0.818434 & 0.539034 & 0.711696 & -0.566396  \\
-0.570235  & 0.796632 & 0.384326 & 0.801647 \\
\hdashline 
          &           & 0.315289 & 0.056348 \\
          &           & 0.136765 & 0.119037 
\end{array}
\]
Whatever the dominant characters of the eigenvectors and the relative positions  of the corresponding eigenvalues in the spectrum, the relative phases of the 3rd and 4th components force us to accept the natural correlation between the 3rd and 4th eigenvectors of ${\bf H}^4$ with the 4th and 3rd eigenvectors of ${\bf H}^6$. This is the signature of a root-flipping that we wanted to illustrate through a simple numerical example.  
\clearpage

\end{document}